\documentstyle[epsfig]{mn}
\newcommand{\pfrac}[2]{\left(\frac{#1}{#2}\right)}

\def\Mesz{M\'esz\'aros~}
\def\Pacz{Paczy\'nski~}

\def\beq{\begin{equation}}
\def\enq{\end{equation}}
\def\bea{\begin{eqnarray}}
\def\ena{\end{eqnarray}}
\def\bec{\begin{center}}
\def\enc{\end{center}}
\def\etal{{et al.~}}
\def\eps{\epsilon}

\def\mujy{\hbox{$\mu$Jy}}

\def\ergcm2si{\hbox{ergs~cm$^{-2}$s$^{-1}$}}

\title{Temporal variability in early afterglows of short gamma-ray bursts}
\author[]{Zhuo Li$^{1,2}$\thanks{lizhuo@mail.ihep.ac.cn},
                  Z. G. Dai$^1$\thanks{daizigao@public1.ptt.js.cn}
          and T. Lu$^1$\thanks{tlu@nju.edu.cn}\\
\rm $^1$Department of Astronomy, Nanjing University, Nanjing
210093, China\\
\rm $^2$Particle Astrophysics Lab., Institute of High Energy
Physics, Chinese Academy of Sciences, Beijing 100039, China}

\date{({\em MNRAS} accepted)}
\pubyear{2003}

\begin{document}

\maketitle

\begin{abstract}
The shock model has successfully explained the observed behaviors
of afterglows from long gamma-ray bursts (GRBs). Here we use it to
investigate the so-called early afterglows from short GRBs, which
arises from blast waves that are not decelerated considerably by
their surrounding medium. We consider a nearby medium loaded with
$e^{\pm}$ pairs (Beloborodov 2002). The temporal behaviors show
first a soft-to-hard spectral evolution, from the optical to hard
X-ray, and then a usual hard-to-soft evolution after the blast
waves begin to decelerate. The light curves show variability, and
consist of two peaks. The first peak, due to the pair effect, can
be observed in the X-ray, though too faint and too short in the
optical. The second peak will be easily detected by {\it Swift}.
We show that detections of the double-peak structure in the light
curves of early afterglows are very helpful to determine all the
shock parameters of short GRBs, including both the parameters of
the relativistic source and the surroundings. Besides, from the
requirement that the forward-shock emission in short GRBs should
be below the BATSE detection threshold, we give a strong
constraint on the shock model parameters. In particular, the
initial Lorentz factor of the source is limited to be no more than
$\sim 10^3$, and the ambient medium density is inferred to be low,
$n\la 10^{-1}$~cm$^{-3}$.
\end{abstract}

\begin{keywords}
gamma-rays: bursts --- radiation mechanisms: nonthermal
--- relativity
\end{keywords}

\section{Introduction}
It is recognized that  gamma-ray bursts (GRBs) may be divided into
at least two classes: one third of the bursts with short duration
($\la 2$~s) and hard spectra, and the other two third with long
duration ($\ga 2$~s) and soft spectra (Kouveliotou \etal 1993;
Dezalay \etal 1996; Paciesas \etal 2003). The detections of
afterglows from long/soft GRBs and then their redshift
measurements have revealed their cosmological origin (see van
Paradijs \etal 2000 for a review). Their afterglows are widely
believed to come from a blast wave driven by a relativistic ejecta
into an ambient medium (see reviews of Cheng \& Lu [2001] and
\Mesz [2002]). Unfortunately, it is impossible so far for
observations to systematically follow short GRBs at longer
wavelengths. The effort of searching transient afterglow emission
from short/hard GRB usually yields only some upper limits (e.g.
Kehoe \etal 2001; Hurly \etal 2002; Gorosabel \etal 2002; Klotz,
Bo\"er \& Atteia 2002).
The difficulty for detection of short GRB afterglow is mainly due
to the poor prompt localization by current satellites for these
bursts. This problem is waiting for the upcoming {\it Swift}
satellite to resolve. Lazzati, Remirez-Ruiz \& Ghisellini (2001)
report the discovery of a $\sim30$~s delayed, transient and fading
hard X-ray emission in the BATSE light curves of a sample of short
GRBs, the soft power-law spectrum and the time-evolution are
consistent with predicted by the afterglow model.

Based on the widely accepted blast wave model, Panaitescu \etal
(2001) studied the long-term afterglows of short GRBs coming from
the blast waves. In this paper, we focus on the investigation of
early afterglow emission, which arises from the blast wave before
it transits to the self-similar evolution in the ambient medium
(Blandford \& McKee 1976). We consider pair loading in the
external medium, which is caused by the collision between the
outgoing gamma-rays and the scattered photons off the external
medium (Madau \& Thompson 2000; Thompson \& Madau 2000; Dermer \&
B\"ottcher 2000; Madau, Blandford \& Rees 2000; \Mesz,
Ramirez-Ruiz \& Rees 2001; Beloborodov 2002; Ramirez-Ruiz,
MacFadyen \& Lazzati 2002). The pairs will affect the behavior of
early afterglows. As the short GRBs have $\sim 20$ times less
fluence than long GRBs (Mukherjee \etal 1998), the kinetic energy
of short GRBs must be $\sim 20$ times less than long GRBs too,
provided that the efficiencies for producing gamma-rays are the
same for both classes (Panaitescu \& Kumar 2001). We take the
typical kinetic energy of short GRBs as $10^{52}$~ergs here.
Furthermore we  assume that the shocks in short GRBs have
parameters similar to those of long GRBs, except for the ambient
density which is believed to be lower if short GRBs originate from
the compact binary mergers (Eichler \etal 1989; Narayan, \Pacz \&
Piran 1992). Later on we will show that low density for a short
GRB is required (eq. (\ref{constraint}) and discussions below). In
section 2 we discuss the hydrodynamics of short GRBs, and in
section 3, pair loading in the external medium. An early afterglow
from the blast wave is derived in section 4. Section 5 gives
conclusions and observational implications.

\section{Hydrodynamics of short GRBs}
A GRB itself is believed to come from internal shocks which are
due to different Lorentz factors of shells within the ejecta (Rees
\& \Mesz 1994). After producing GRB the ejecta cools down rapidly
and may be considered as a cold shell. The interaction between the
outgoing shell and ambient medium leads to two shocks: a forward
shock propagating into the medium, a reverse shock sweeping up the
ejecta matter, and a contact discontinuity separating the shocked
ejecta matter and the shocked medium. So the kinetic energy of the
ejecta can be dissipated into the internal energy of the medium by
the forward shock and into the internal energy of the ejecta
matter by the reverse shock. According to Sari (1997), there are
two time scales. One is relevant to the forward shock, at which
the shell reaches an deceleration radius where the shell has given
the medium an energy comparable to its initial energy,
\begin{equation}\label{tdec}
t_{\rm
dec}=45E_{k,52}^{1/3}\eta_{300}^{-8/3}n_{-2}^{-1/3}\pfrac{1+z}2~{\rm
s},
\end{equation}
where $E_k=10^{52}E_{k,52}$~ergs and $\eta=300\eta_{300}$ are the
fireball kinetic energy and initial Lorentz factor,
$n=0.01n_{-2}$~cm$^{-3}$ is the particle density of the ambient
medium, and $z$ is the source's redshift. The other is relevant to
the reverse shock, at which the reverse shock accelerates to
become relativistic. The ratio between the two time scales is
defined as
\begin{equation}\label{xi}
\xi=12E_{k,52}^{1/6}\pfrac\Delta{3\times10^9{\rm
cm}}^{-1/2}\eta_{300}^{-4/3}n_{-2}^{-1/6},
\end{equation}
where $\Delta$ is the shell width (in observer frame) of the
ejecta. In the internal-shock model the shell width is
$\Delta=cT=3\times10^9T_{-1}$~cm, with $T=0.1T_{-1}$ s the
duration of GRB. Eq. (\ref{xi}) shows that $\xi$ is not sensitive
to $E_k$ and $n$, and only somewhat dependent on $\eta$ which is
not accepted to be quite larger than $10^3$ (implied from
eq.[\ref{constraint}] below, and also implied from other aspects
of GRBs, e.g., Lazzaiti, Ghisellini \& Celotti 1999; Derishev,
Kocharovsky \& Kocharovsky 2001). Thus, for short GRBs, we usually
have $\xi>1$. In this case, the reverse shock is initially
Newtonian and becomes mildly relativistic when it crosses the
shell at $t_{\rm dec}$. Consequently the shocked medium has most
of the initial energy, and the forward shock goes into the
self-similar Blandford-McKee (1976) evolution.

\section{Pair loading in GRB medium}
The GRB itself from internal shocks is emitted early, preceding
the development of the blast wave. The gamma-ray front interacts
with the ambient medium, leading to two processes: Compton
scattering and $\gamma-\gamma$ absorption of the scattered
photons. As a result the medium is loaded with $e^{\pm}$ pairs
within a loading radius $R_{\rm
load}=5\times10^{15}E_{\gamma,52}^{1/2}$~cm, with
$E_\gamma=10^{52}E_{\gamma,52}$~ergs the isotropically explosive
energy in gamma-rays (Beloborodov 2002). Approximately $10^3$
pairs per ambient electron can be created when conditions are
right, but usually it is much less, $f_0\equiv N_{\pm}/N(R_{\rm
load})=10^2f_{0,2}$ (Beloborodov 2002). Therefore, the mass of
$e^{\pm}$ pairs ahead of the blast wave is neglected,
$f_0<m_p/m_e$, it dose not affect the dynamics of the blast wave.
Besides, the pairs may be pre-accelerated by the gamma-ray front,
but the pair energy does not exceeds the ejecta kinetic energy.
Provided that the medium density is low and the deceleration
occurs outside the pre-accelerated radius $R_{\rm acc}$
(Beloborodov 2002) which is smaller than $R_{\rm load}$, the
deceleration time (eq.[\ref{tdec}]) will be not affected (Note
however that as shown by Beloborodov [2002], for dense enough
medium $t_{\rm dec}$ changes whenever deceleration occurs in a
pre-accelerated medium, i.e., $R_{\rm dec}<R_{\rm acc}$).
Typically, the deceleration time is longer than the one at which
the blast wave approaches $R_{\rm load}$,
\begin{equation}\label{tload}
t_{\rm load}=\frac{R_{\rm
load}(1+z)}{2\eta^2c}=1.7E_{\gamma,52}^{1/2}\eta_{300}^{-2}\pfrac{1+z}2~{\rm
s},
\end{equation}
and the one at which the blast wave crosses a radius,
$R_f=f_0^{1/3}R_{\rm load}$ where the number ratio ($f$) of the
pair to ambient electron number drops to $f=1$,
\begin{equation}\label{tf}
t_f=7.9E_{\gamma,52}^{1/2}\eta_{300}^{-2}f_{0,2}^{1/3}\pfrac{1+z}2~{\rm
s}.
\end{equation}
Thus, for short GRBs we have the order, $t_{\rm load}<t_f<t_{\rm
dec}$.

Here, when introducing radius $R_f$, we have assumed mixing of
particles in the blast wave that allows the newly added post-shock
particles to share energy with earlier injected pairs. Before the
reverse shock crosses the ejecta and vanishes at $t_{\rm dec}$,
the existence of the contact discontinuity prevents the earlier
pairs from being far downstream the forward shock front. The total
shocked mediums are compressed between the contact discontinuity
and the forward shock front. Furthermore, the coupling of leptons
with baryons may take place in presence of even weak magnetic
fields (e.g. Madau \& Thompson, 2000; Meszaros, Ramirez-Ruiz \&
Rees 2001), therefore the total particles are possible to be
mixing, allowing continuous transmission of energy from baryons to
leptons. We will take the mixing hypothesis in the following.

\section{Early afterglows of short GRBs}
We now derive the temporal property of early afterglows from
forward shocks of short GRBs. We  consider the source as an
isotropic explosion even though it may have jet geometry, because
the jet effect is not important at early times when the jet open
angle is larger than $\sim1/\eta$.

\subsection{Phase $t_{\rm load}<t<t_f$}
We begin with the blast wave having swept up all the produced
pairs at $R_{\rm load}$.  The pairs will modify the usual property
of the afterglow, since the same energy will be shared by much
more leptons. Furthermore  the pairs will increase the radiation
efficiency significantly. With the mixing hypothesis, the
comoving-frame random lepton Lorentz factor is
\begin{equation}
\gamma_m=\frac{m_p}{(1+f)m_e}\eps_e\eta.
\end{equation}
Here, $f\equiv N_{\pm}/N_e$, and the energy density in leptons and
magnetic field ${B^2\over4\pi}$ behind the shock are usually
parameterized by the fractions $\eps_e=0.1\eps_{e,-1}$ and
$\eps_B=0.01\eps_{B,-2}$ of the total internal energy density
($\eta^2nm_pc^2$), respectively. For $f>1$ at $t_{\rm
load}<t<t_f$, this Lorentz factor is a factor $(1+f)\approx f$
lower than usual case, and the corresponding  synchrotron
frequency is therefore
\begin{equation}\label{numload}
\nu_m=1.8\times10^{14}\eps_{e,-1}^2\eps_{B,-2}^{1/2}\eta_{300}^4n_{-2}^{1/2}f_2^{-2}\pfrac{1+z}2^{-1}~{\rm
Hz},
\end{equation}
which is in the optical band if  $f=f_0=10^2f_{0,2}$ at $t_{\rm
load}$, as opposed to the hard X-ray of the usual case. Now the
pair number dominates the ambient electron's number, the total
lepton number is $N_{\rm lep.}\simeq{4\over3}\pi R_{\rm
load}^3nf_0$. The peak spectral power (in comoving frame) per
lepton is
$P_{\nu,\max}=1.4\times10^{-22}B$~ergs\,s$^{-1}$Hz$^{-1}$. We then
have the afterglow peak flux
\begin{eqnarray}
F_p&=&N_{\rm lep.}\eta P_{\nu,\max}\frac{(1+z)}{4\pi d_l^2}\nonumber\\
&=&3.2\eps_{B,-2}^{1/2}E_{k,52}^{3/2}\eta_{300}^2n_{-2}^{3/2}f_{0,2}d_{l,28}^{-2}\pfrac{1+z}2\mujy
\end{eqnarray}
where $d_l=10^{28}d_{l,28}$ is the GRB's luminosity distance. To
calculate the synchrotron spectrum, we still need to know the
cooling frequency that is corresponding to those leptons which
cool by synchrotron/inverse-Compton radiation in a dynamical time
$t$, i.e.
\begin{equation}\label{nuc}
\nu_c=2.5\times10^{20}\eps_{B,-2}^{-3/2}\eta_{300}^{-4}n_{-2}^{-3/2}t^{-2}\pfrac{1+Y}2^{-2}\pfrac{1+z}2~{\rm
Hz},
\end{equation}
where $Y$ is the Compton parameter. According to Panaitescu \&
Kumar (2000),
$Y={1\over2}\{[{5\over6}(\eps_e/\eps_B)+1]^{1/2}-1\}\approx1$. Now
for observer's time $t=t_{\rm load}$,
\begin{equation}
\nu_c(t_{\rm
load})=3.4\times10^{20}\eps_{B,-2}^{-3/2}E_{\gamma,52}^{-1}n_{-2}^{-3/2}\pfrac{1+z}2^{-1}~{\rm
Hz}.
\end{equation}
The synchrotron spectrum from leptons distributed as $dN_{\rm
lep.}/d\gamma_e\propto\gamma_e^{-p}~(\gamma_e>\gamma_m)$ is a
broken power-law with break frequencies  $\nu_m$ and $\nu_c$:
$F_\nu\propto\nu^{1/3}$ at $\nu<\nu_p\equiv\min(\nu_m,\nu_c)$;
$F_\nu\propto\nu^{-1/2}$ for $\nu_c<\nu<\nu_m$ or
$F_\nu\propto\nu^{-(p-1)/2}$ for $\nu_m<\nu<\nu_c$; and
$F_\nu\propto\nu^{-p/2}$ at $\nu>\max(\nu_m,\nu_c)$ (Sari, Piran
\& Narayan 1998). Here we neglect the synchrotron self-absorption
which is only important at longer wavelengths, e.g., radio or IR.

Since $f\propto N_e^{-1}\propto R^{-3}\propto t^{-3}$,
eq.(\ref{numload}) implies that the peak frequency rapidly
increases, as $\nu_m\propto t^6$, from the optical to the hard
X-ray band eventually (eq.[\ref{numf}]). Thus, we have the scaling
laws for $t_{\rm load}<t<t_f$,
\begin{equation}
F_p={\rm const.},~~\nu_m\propto t^6,~~\nu_c\propto t^{-2}~~(t_{\rm
load}<t<t_f).
\end{equation}
The afterglow shows a soft-to-hard spectral evolution during this
phase. Observed at a fixed frequency, $\nu_{\rm ob}$, between the
optical and hard X-ray, the light curve will show a rapidly
increase, $F_\nu\propto t^{3(p-1)}$, and then a sharp decreasing,
$F_\nu\propto t^{-2}$, after $\nu_m$ crosses $\nu_{\rm ob}$ at
\begin{equation}\label{tpk}
t_{pk}=4.9\frac{\nu^{1/6}_{{\rm
ob},17}E_{\gamma,52}^{1/2}f_{0,2}^{1/3}}{\eps_{e,-1}^{1/3}(\eps_{B,-2}n_{-2})^{1/12}\eta^{8/3}_{300}}\pfrac{1+z}2^{7/6}{\rm
s}.
\end{equation}

\subsection{Phase $t_f<t<t_{\rm dec}$}
outside $R_f$, we have $f<1$, implying that pair effect is
negligible. The afterglow property then approaches the usual case,
where
\begin{equation}
F_p\propto N_e\propto t^3,~~\nu_m={\rm const.},~~\nu_c\propto
t^{-2}~~(t_f<t<t_{\rm dec}).
\end{equation}
In details,
\begin{equation}\label{numf}
\nu_m=1.8\times10^{18}\eps_{e,-1}^2\eps_{B,-2}^{1/2}E_{k,52}\eta_{300}^4n_{-2}^{1/2}\pfrac{1+z}2^{-1}~{\rm
Hz}
\end{equation}
is a constant and should be in the hard X-ray band. The  cooling
frequency continues to decrease (eq.[\ref{nuc}]) to
\begin{equation}
\nu_c(t_{\rm dec})
=5.0\times10^{17}\eps_{B,-2}^{-3/2}E_{k,52}^{-2/3}\eta_{300}^{4/3}n_{-2}^{-5/6}\pfrac{1+z}2^{-1}~{\rm
Hz}
\end{equation}
at $t=t_{\rm dec}$. Note that $\nu_m>\nu_c(t_{\rm dec})$, implying
that $\nu_c$ has crossed $\nu_m$ at a certain moment $t_{cm}$
after which the spectrum becomes peaking at $\nu_c$, which is in
X-rays. Due to ambient electrons picked up, the peak flux
increases rapidly to
\begin{equation}\label{fpdec}
F_p(t_{\rm
dec})=580\eps_{B,-2}^{1/2}E_{k,52}n_{-2}^{1/2}d_{l,28}^{-2}\pfrac{1+z}2\mujy
\end{equation}
at $t=t_{\rm dec}$. If observing at a fixed sub-keV frequency, we
can see in this phase the light curve climbing up again.

Now a constraint on short GRBs arises from the requirement that
the flux in sub-MeV should not exceed the BATSE detection
threshold. Otherwise, as $t_{\rm dec}>2$~s, the burst is not short
any more. With $n=0.01$ and other parameters in their typical
value, we obtain the flux given by
\begin{eqnarray}
\Phi({\rm MeV})\simeq& 2\nu_mF_{\nu_m}=2(\nu_m\nu_c)^{1/2}F_p=1.1\times10^{-8}\nonumber\\
&\times\eps_{e,-1}E_{k,52}^{2/3}\eta_{300}^{8/3}n_{-2}^{1/3}d_{l,28}^{-2}
\hbox{ergs~cm$^{-2}$s$^{-1}$}.
\end{eqnarray}
We set that the  BATSE threshold is
$1\times10^{-8}$~ergs~cm$^{-2}$s$^{-1}$, leading to a constraint
on the ``short GRB'' parameters of
\begin{eqnarray}\label{constraint}
\eps_{e,-1}E_{k,52}^{2/3}\eta_{300}^{8/3}n_{-2}^{1/3}d_{l,28}^{-2}
<1.
\end{eqnarray}
Note that the most stringent constraint is on $\eta$, which is not
allowed to be too large, i.e. $\eta\la 10^3$. A lower limit to
$\eta$ arises from the requirement that during the prompt sub-MeV
burst, the optical depth due to scattering off fireball electrons,
$\tau_b=\sigma_T N_b/4\pi R_\gamma^2$, should be less than unity
(Rees \& \Mesz 1994), with $N_b=E_k/\eta m_pc^2$ the fireball
baryon number, $R_\gamma\leq\eta^2 c\delta t$ the radius at which
the fireball kinetic energy dissipated to gamma rays, and $\delta
t$ the shortest time scale of rapid variability in the GRB
profile. This leads to $\eta>330E_{k,52}^{1/5}\delta
t_{-2}^{-2/5}$, thus the $\eta$ value taken in eq.
(\ref{constraint}) is to the lower limit. If the other parameters
are fixed to their typical values, the ambient density for short
GRBs is limited to $n\la 0.01$~cm$^{-3}$ (eq. [\ref{constraint}]),
consistent with the clean-environment hypothesis to short GRB
models of compact binary systems, e.g. Eichler \etal (1989);
Narayan, \Pacz \& Piran (1992).

\subsection{Phase $t>t_{\rm dec}$}
In this phase, the blast wave begins to decelerate considerably.
If the electrons obtain a significant fraction of total energy,
$\eps_e\sim1$, the blast wave will evolve in the radiative regime,
since all the electrons are fast cooling, with $\nu_c<\nu_m$. The
light curve is somewhat complicated in this case with light-curve
index related to $\eps_e$ (B\"ottcher \& Dermer 2000; Li, Dai \&
Lu 2002). For the typical value $\eps_e=0.1$, we can safely
consider a adiabatic blast wave, so the well know scaling laws
are:
\begin{equation}
F_p={\rm const.},~~\nu_m\propto  t^{-3/2},~~\nu_c\propto
t^{-1/2}~~(t>t_{\rm dec}),
\end{equation}
where $F_p$ is given by eq.(\ref{fpdec}). The afterglow spectrum
shows the usual hard-to-soft evolution after $t_{\rm dec}$.
Observed at a certain frequency $\nu_{\rm ob}$ between the optical
and keV band, when $\nu_m$ or $\nu_c$ crosses $\nu_{\rm ob}$,
whichever the first, the observed flux reaches a peak with $F_{\rm
ob}=F_p\simeq580\mu{\rm Jy}$. It is  a magnitude $\simeq15.6$ if
observed in the optical. Thus, there is another peak in the light
curve other than the first one in the phase $t<t_f$. Furthermore,
this second peak is much stronger than the first one. Lazzati,
Remirez-Ruiz \& Ghisellini (2001) claim to have detected such a
delayed hard X-ray peak.

In Fig. 1 we show the light curves at two bands, the optical and
the X-ray, also labelled in this figure are the characteristic
times and the light-curve scaling laws.

\begin{figure}
\centerline{\hbox{\psfig{figure=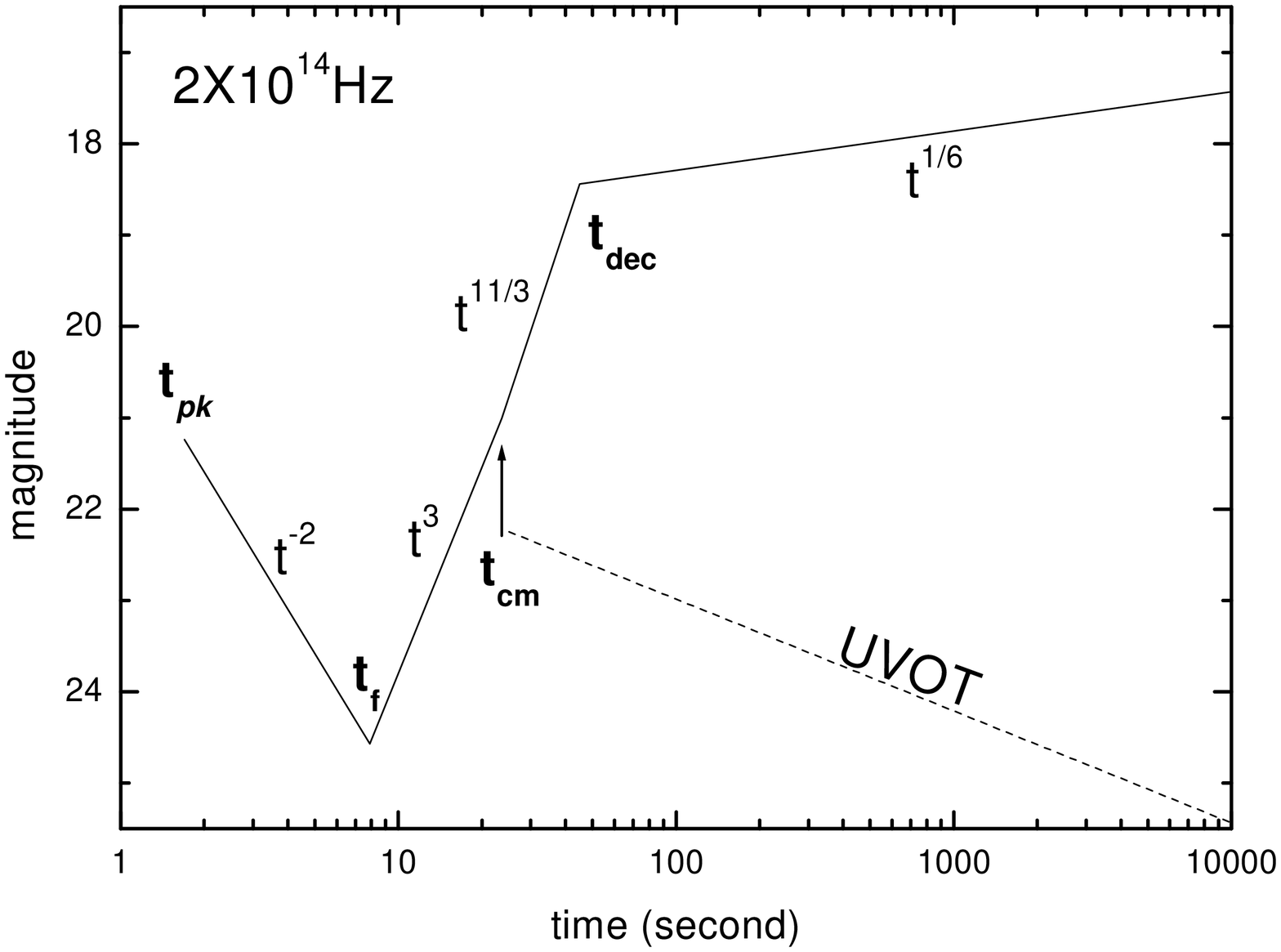,width=\columnwidth,angle=0}}}
\centerline{\hbox{\psfig{figure=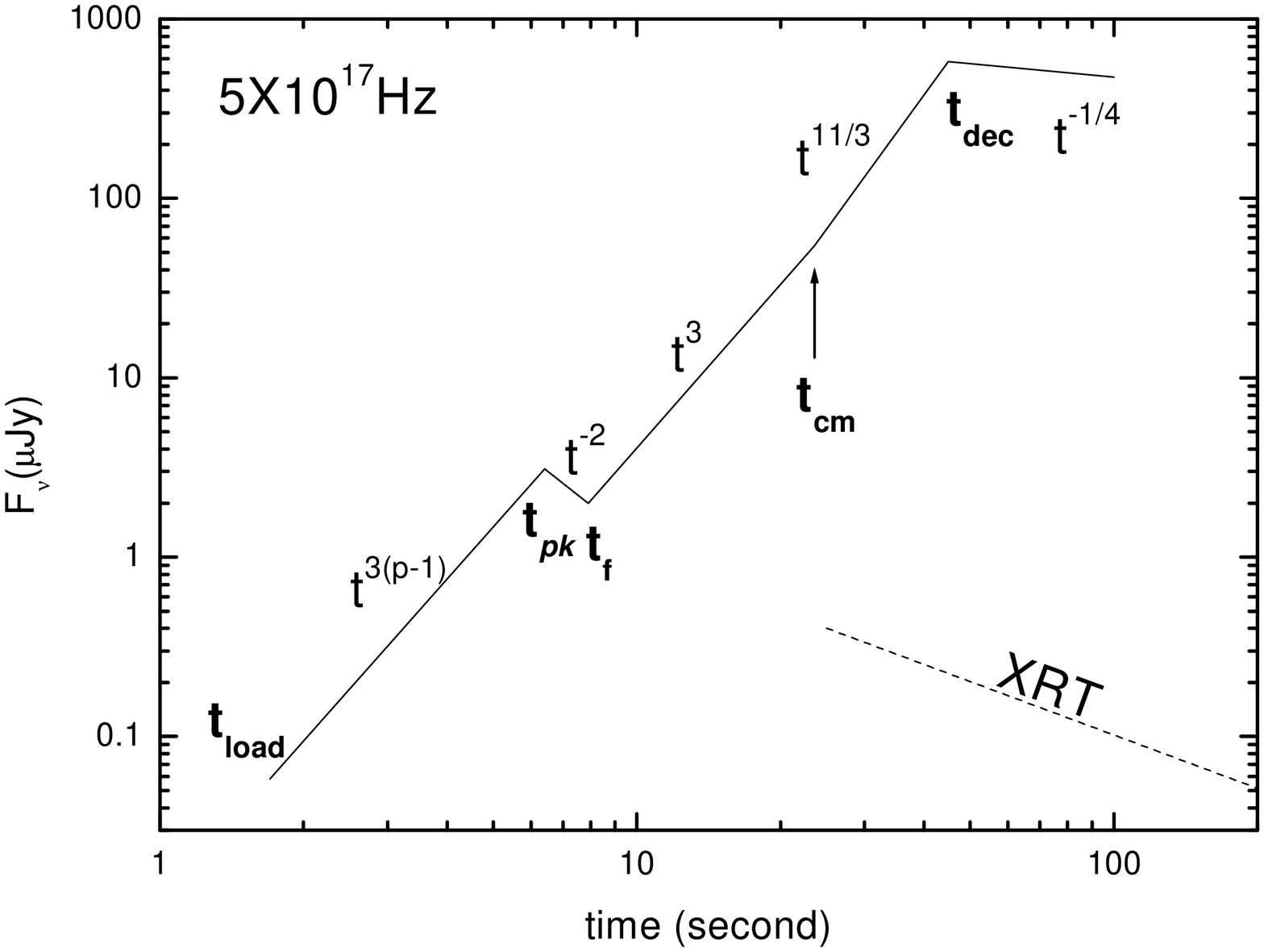,width=\columnwidth,angle=0}}}
\caption {Example of early afterglows of from short GRBs at two
fixed frequencies, $\nu=2\times10^{14}$ ({\it upper frame}) and
$5\times10^{17}$ ({\it bottom frame}) Hz. The parameter values
taken to calculate the light curves are:
$E_\gamma=E_k=10^{52}$~ergs, $n=0.01$~cm$^{-3}$, $f_0=10^2$, $p=2$
and the others equal to typical values of long GRBs (see details
in the text). The characteristic times and the scaling laws of
fluxes with time are marked. The dashed lines show the sensitivity
of {\it Swift} instruments, the X-ray (XRT) and UV optical (UVOT)
Telescopes. }
\end{figure}

\section{Conclusion and discussion}
Based on the shock model which has been essentially successful to
explain long GRB afterglows, we here derive the light curves of
short GRB afterglows in the early phase when the blast wave is not
decelerated by the ambient medium considerably. The reverse-shock
emission has been ignored at the beginning since it is always
Newtonian initially for short GRBs. We consider the pair-loading
effects on the emission. The spectrum shows rapid soft-to-hard
evolution at the first several seconds ($t<t_f$), and then a usual
hard-to-soft evolution after several tens of seconds ($t>t_{\rm
dec}$). Simultaneously,  there are two peaks appearing in the
light curves in the optical to hard X-ray range. The first ``pair
peak" will appear at the optical, but it is too faint (mag
$\sim21$) and too short ($\sim 2$~s) to be detected by any current
and upcoming instrument. But the double-peak structure in the
light curve is expected to be observed at X-ray band: the first
peak at $t_{pk}$ and the second peak at around $t_{\rm dec}$. It
took only 20 to 70 s for {\it Swift} to point its Narrow Fields
instruments, consisting of X-ray and UV optical Telescopes, to the
GRB direction, short GRBs will be easily detected before the
second peak (Fig. 1). The recently proposed micro-satellite {\it
ECLAIR} (Barret 2003) is even expected capable of detecting the
first peak.

Though the reverse shock becomes mildly relativistic finally at
$t_{\rm dec}$, we have neglected its emission here, which is
mainly in the soft band, say, the optical. If we consider further
the effect of pair-loading in the fireball which due to
$\gamma-\gamma$ absorption of the prompt burst in the fireball, it
would be in much softer band such as in the IR. Because the same
energy may be shared by more produced $e^{\pm}$ pairs,  the
lower-energy leptons would radiate at softer frequency. So the
reverse-shock emission will not affect the X-ray light-curve,
though may affect the optical one.

The blast wave emission in sub-MeV must be under the BATSE
detection threshold for short GRBs. Provided that the energy is
$E_k=10^{52}$~ergs, and $\eps_{e,-1}=d_{l,28}=1$ similar to
typical values of long GRBs, we find a constraint on initial
Lorentz factor and ambient density of
$\eta^{8/3}_{300}n^{1/3}_{-2}<1$ (cf.eq.[\ref{constraint}]), and
that the Lorentz factor of short GRBs is not allowed to be large,
i.e., $\eta<10^3$. This also limit the ambient density to $n\la
0.1~{\rm cm}^{-3}$, which is consistent with upper limit on
late-time short GRB afterglows. So far, the best constraint on
short GRB afterglows comes from the observation of short/hard GRB
020531, which yields the limiting magnitudes in R band: 18.5 at 88
min and 25.2 at 2.97 d (Klotz \etal 2002). Under a standard
afterglow model, these data do not allow for a dense medium, i.e.
$n\la 0.1~{\rm cm}^{-3}$ (see Fig. 2 in Panaitescu \etal 2001).
Low densities favor the GRB model related to compact object
mergers (Eichler \etal 1989; Narayan, \Pacz \& Piran 1992) in
galactic haloes or in the intergalactic medium.

At $t_{\rm load}(>2$~s) the optical photons in the pulse may be
up-scattered to MeV  by synchrotron self-Compton process. But the
flux is of orders lower than BATSE detection threshold, and is
unable to change the short-duration property of short GRBs.

Unlike the long GRBs which may overlap the early afterglows and
lead to complication, the short GRBs stop abruptly. And  due to
lower ambient density the blast waves take longer time to begin
decelerating considerably, so their early afterglows are easy to
be observed. If detected and confirmed, the double-peak structure
in early afterglows has an important indication for short GRBs ---
with redshift having been measured, we can determine the most
important parameter $\eta$ from eqs.(\ref{tf}) and (\ref{tpk})
(Beloborodov 2002), and then we can further use the value of
$\eta$ to constraint the other parameters, $E_k$ and $n$, by
eq.(\ref{tdec}). So an observation of early afterglows provides
important constraints on the short GRB parameters, related both to
the relativistic flow and to the surroundings.

\section*{Acknowledgments}
We would like to thank the referee for valuable comments. Z. Li
thanks X.Y. Wang for valuable discussions and R.F. Shen for
careful reading. This work was supported by the National Natural
Science Foundation of China, the National 973 Project (NKBRSF
G19990754) and the Special Funds for Major State Basic Research
Projects.


\end{document}